\documentclass[twoside,twocolumn,10pt]{article}

\usepackage{wscg}           
\RequirePackage{ifpdf}
\ifpdf
 \RequirePackage[pdftex]{graphicx}
 \RequirePackage[pdftex]{color}
\else
 \RequirePackage[dvips,draft]{graphicx}
 \RequirePackage[dvips]{color}
\fi
\usepackage{amsmath}  

\usepackage{nopageno}       

\title{PCA for Enhanced Cross-Dataset Generalizability in Breast Ultrasound Tumor Segmentation}

\author{
\parbox{0.3\textwidth}{\centering
Christian Schmidt\\[1mm]
Westf\"alische Hochschule University of Applied Sciences\\
Neidenburger Strasse 43\\
45897 Gelsenkirchen, Germany\\[1mm]
christian.schmidt@w-hs.de
}
\hspace{0.05\textwidth}
\parbox{0.3\textwidth}{\centering
Heinrich Martin Overhoff\\[1mm]
Westf\"alische Hochschule University of Applied Sciences\\
Neidenburger Strasse 43\\
45897 Gelsenkirchen, Germany\\[1mm]
martin.overhoff@w-hs.de
}
}

\usepackage{url}
\makeatletter
\def\Uslash{\mathbin{\mathchar`\/}\@ifnextchar{/}{\kern-.15em}{}}
\g@addto@macro\UrlSpecials{\do \/ {\Uslash}}
\def\Ucolon{\mathbin{\mathchar`:}\@ifnextchar{/}{\kern-.1em}{}}
\g@addto@macro\UrlSpecials{\do : {\Ucolon}}
\makeatother

\begin{document}

\twocolumn[{\csname @twocolumnfalse\endcsname

\maketitle  

\begin{abstract}
\noindent
In medical image segmentation, limited external validity remains a critical obstacle when models are deployed across unseen datasets, an issue particularly pronounced in the ultrasound image domain. Existing solutions-such as domain adaptation and GAN-based style transfer-while promising, often fall short in the medical domain where datasets are typically small and diverse. This paper presents a novel application of principal component analysis (PCA) to address this limitation. PCA preprocessing reduces noise and emphasizes essential features by retaining approximately 90\% of the dataset variance. We evaluate our approach across six diverse breast tumor ultrasound datasets comprising 3,983 B-mode images and corresponding expert tumor segmentation masks. For each dataset, a corresponding dimensionality reduced PCA-dataset is created and U-Net-based segmentation models are trained on each of the twelve datasets. Each model trained on an original dataset was inferenced on the remaining five out-of-domain original datasets (baseline results), while each model trained on a PCA dataset was inferenced on five out-of-domain PCA datasets. Our experimental results indicate that using PCA reconstructed datasets, instead of original images, improves the model's recall and Dice scores, particularly for model-dataset pairs where baseline performance was lowest, achieving statistically significant gains in recall (0.57 $\pm$ 0.07 vs. 0.70 $\pm$ 0.05, $p = 0.0004$) and Dice scores (0.50 $\pm$ 0.06 vs. 0.58 $\pm$ 0.06, $p = 0.03$). Our method reduced the decline in recall values due to external validation by $33\%$. These findings underscore the potential of PCA reconstruction as a safeguard to mitigate declines in segmentation performance, especially in challenging cases, with implications for enhancing external validity in real-world medical applications. Future studies are proposed to optimize PCA configurations for diverse imaging datasets and exploring integration with existing external validation methods.

\end{abstract}

\subsection*{Keywords}
Breast tumor segmentation, domain generalization, ultrasound imaging, neural networks, principal component analysis

\vspace*{1.0\baselineskip}
}]


\setlength{\tabcolsep}{2.3 pt}
\begin{table*}[]
	\centering
	\begin{tabular}{|p{2cm}|p{2,1cm}|p{6cm}|p{1,3cm}|} 
		\hline
		\textbf{Name} & \textbf{No. of cases} & \textbf{Ultrasound machine and transducer} & \textbf{Release year} \\ \hline
		Ardakani & 109 benign \newline 123 malignant & AirPlorer Ultimate \newline linear transducer (5-18 MHz)
		& 2023 \\ \hline
		BrEaST & 154 benign \newline 98 malignant & Hitachi ARIETTA 70 \newline linear array transducer L441 (2-12 MHz) & 2024 \\ \hline
		BUS\_UC & 358 benign \newline 453 malignant & Hitachi \newline various transducers & 2023 \\ \hline
		BUSBRA & 722 benign \newline 342 malignant & GE Logiq 5, GE Logiq 7, Toshiba Aplio 300, GE U-Systems (all 10-14 MHz) & 2023\\ \hline
		BUSI & 487 benign \newline 210 malignant & LOGIQ E9 \newline ML6-15-D Matrix linear probe (1-5 MHz)
		& 2020\\ \hline
		BUSI\_WHU & 927 total & not specified & 2023\\ \hline
	\end{tabular}
	\caption{Overview of the datasets used in this study, showing dataset name, number of images in the dataset, model of ultrasound machine and transducer as well as the year of dataset release.}
\end{table*}

\section{Introduction}

\copyrightspace

Breast cancer surgeries face significant challenges, with reoperation rates reaching 15-20\% due to insufficient excision precision~\cite{Kim}. Improved accuracy in tumor segmentation for ultrasound imaging could help mitigate this issue by supporting more precise, intraoperative navigation. Consequently, achieving robust performance in fully automated breast tumor segmentation is essential for enhancing surgical outcomes. 
One of the key challenges in machine-learning-based image segmentation is poor external validity, where models trained on one dataset often perform poorly when applied to other, unseen datasets. Previous studies~\cite{Yu} have already underscored this problem, highlighting its detrimental impact on the generalizability in medical applications. This limitation undermines the reliability of automated systems in real-world medical settings, where data variability is high. Addressing this issue is critical to ensuring that segmentation models can generalize effectively across diverse patient populations and imaging conditions. 
Principal Component Analysis (PCA) reduces the dimensionality of the data, preserving essential features while minimizing variability, thus improving the generalizability of segmentation models across datasets. When training a segmentation model on ultrasound images from two devices, without PCA, the model might overfit to device-specific noise (e.g., machine artifacts or probe settings). With PCA, the preprocessing step removes such device-specific details, resulting in training data that emphasizes tumor structures common across devices. This process is similar to having a more diverse dataset that naturally includes a variety of imaging conditions. In essence, PCA preprocessing doesn't add new diversity but reduces the dataset-specific biases that could hinder generalization, making the data behave as if it were drawn from a broader, more uniform distribution. Therefore, our work proposes to leverage PCA as a preprocessing step to improve external validity in breast tumor segmentation. 
\section{Related Work}
External validity in medical image segmentation has been extensively studied, with various methods proposed to address the challenge of domain shifts. Domain adaptation techniques~\cite{Guan}, for example, aim to align feature distributions between training and test datasets, often using adversarial training or feature space regularization. While effective, these methods require access to target domain data during training, which may not always be feasible in medical applications.

GAN-based style transfer~\cite{Liu} has also been explored as a means to harmonize imaging styles between datasets. These approaches generate synthetic images that mimic the target domain's style, but their reliance on large datasets and computational resources can limit their applicability in the medical field. Similarly, synthetic data generation~\cite{Stojanovski} using diffusion models has shown promise for augmenting small datasets, yet such methods are often constrained by the quality and diversity of the generated data.

Another common strategy is transfer learning, where models pre-trained on large-scale datasets like ImageNet are fine-tuned for medical tasks. However, the structural differences between natural and medical images, as well as the large number of parameters in pre-trained models, often result in suboptimal performance~\cite{Guan}.

In contrast to these approaches, PCA offers a computationally efficient alternative by reducing dataset-specific noise and emphasizing shared features across datasets. PCA is preferable to style transfer or GAN-based approaches for small medical datasets because it introduces less risk of overfitting and hallucinated features, ensuring that the transformed images remain closely tied to the original data distribution. While PCA has been widely used for dimensionality reduction in medical imaging, its application to enhance external validity in segmentation tasks remains underexplored. This study seeks to fill this gap by evaluating the effectiveness of PCA preprocessing in improving cross-dataset segmentation performance.

\begin{figure*}[]
	\centering
	\includegraphics[width=\textwidth]{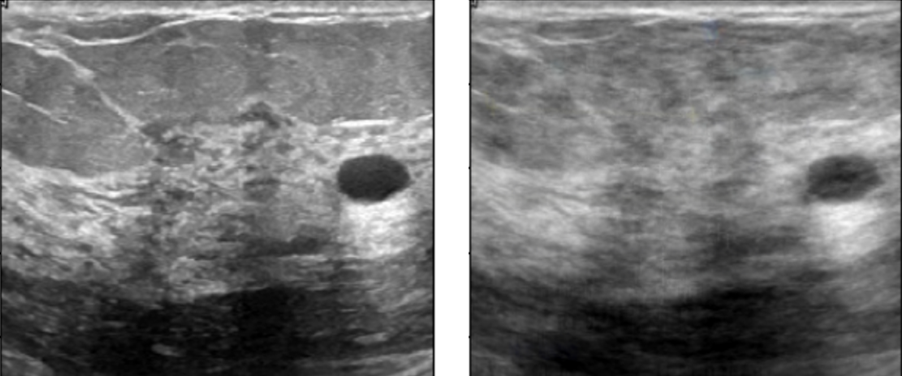} 
	\caption{Comparison of breast tumor image pair before (left) and after (right) PCA reconstruction, showing noise reduction and background smoothing to enhance generalizability across datasets.}
	\label{fig:origpca}
\end{figure*}

\section{Material and Methods}
\subsection{Datasets}
In recent years, the availability of more public breast tumor ultrasound datasets made it possible to conduct research focused on enhancing external validity in segmentation models. For our research, we used six datasets: Ardakani et al.~\cite{Ardakani}, BrEaST~\cite{Breast}, BUS\_UC~\cite{BUSUC}, BUSI~\cite{BUSI}, BUSBRA~\cite{BUSBRA}, and BUSI\_WHU~\cite{BUSWHU}. These datasets encompass a total of 4,116 ultrasound images taken in different settings with various ultrasound machines, depicting both malignant and benign breast tumors. As this work focuses on tumor segmentation, we excluded the 133 normal images that did not feature tumors from our dataset. This led to a total of 3,983 images to be included in our experiments. Each dataset is characterized by its image count, year of acquisition, and the ultrasound machine used to record the images (Table 1).

\subsection{Experiment}
Our segmentation model is based on a U-Net architecture with an input size of 256$\times$256$\times$1. The encoder is comprised of four convolutional blocks. Each block consists of two 3$\times$3 convolutional layers with ReLU activation function, followed by a 2$\times$2 max pooling layer. The number of filters doubles with each block, from 32 in the first layer to 256 in the bottleneck. The decoder mirrors the encoder, using 2$\times$2 transposed convolutions for upsampling, and concatenating the corresponding encoder layers to retain spatial information. The final output is produced by a convolutional layer with a sigmoid activation, providing a single-channel segmentation map. 

Each model was trained using a combined loss function that integrates Dice loss and binary cross-entropy with a weight factor $\beta = 0.5$. The combined loss function enhances segmentation by addressing class imbalance and improving boundary precision~\cite{Galdran}. Before model training, the image data was normalized to a range of $[0,1]$. Training was performed with the Adam optimizer, and the model was trained for a specified maximum of 100 epochs with a batch size of 8 images. Early stopping with a patience of 10 epochs was employed to prevent overfitting, while restoring the best model weights. Consistency was a key focus throughout the experiments, as we maintained the same neural network architecture and training procedure, using the same 70/10/20 training/validation/test split to avoid data leakage (scikit-learn random state = 42).

We assume that each of the six datasets represents a sample from the overall population of breast tumor images. However, because these samples differ significantly from one another and do not adequately represent the full population, PCA is applied to harmonize the datasets. PCA is a dimensionality reduction technique that identifies the directions (principal components) along which the variance of the data is maximized. By projecting the data onto these components, PCA transforms high-dimensional data into a lower-dimensional space while retaining as much variance as possible. PCA was employed to decompose each dataset into its principal components, followed by a reconstruction of the original dataset through a linear combination of these principal components. This is given by
\[
\mathbf{X}_{\text{reconstructed}} = \mathbf{\bar{X}} + \sum_{i=1}^{n} \mathbf{z}_i \mathbf{w}_i^T
\]
where \( \mathbf{X}_{\text{reconstructed}} \) is the reconstructed dataset, \( \mathbf{\bar{X}} \) is the mean of the original data, \( \mathbf{z}_i \) represents the score of the \( i \)-th principal component, \( \mathbf{w}_i \) is the \( i \)-th principal component vector, and \( n \) is the number of principal components used.

\setlength{\tabcolsep}{1.4 pt}
\renewcommand{\arraystretch}{1.5}
\begin{table*}[]
	\centering
	
	\begin{tabular}{|c|ccc|ccc|ccc|ccc|ccc|ccc|}
		\hline
		& \multicolumn{3}{c|}{Ardakani} & \multicolumn{3}{c|}{BrEaST} & \multicolumn{3}{c|}{BUS\_UC} & \multicolumn{3}{c|}{BUSBRA} & \multicolumn{3}{c|}{BUSI} & \multicolumn{3}{c|}{BUSI\_WHU} \\
		\cline{2-19}
		& Ori & PCA & Diff & Ori & PCA & Diff & Ori & PCA & Diff & Ori & PCA & Diff & Ori & PCA & Diff & Ori & PCA & Diff \\
		\hline
		Ardakani &0.82 &0.84 &0.02 &0.50 &0.73 &\textbf{0.23} &0.75 &0.77 &0.02 &0.77 &0.76 &-0.01 &0.66 &0.67 &0.01 &0.74 &0.76 &0.02 \\
		BrEaST  &0.76 & 0.70 &\textbf{-0.06} &0.71 &0.73 &0.02 &0.69 &0.71 &0.02 &0.70&0.72 &0.02 &0.57 &0.59 &0.02 &0.73 &0.70 &-0.03 \\
		BUS\_UC  &0.88 &0.87 &-0.01 &0.65 &0.77 &\textbf{0.12} &0.88 &0.91 &0.03 &0.88 &0.90 &0.02 &0.74 &0.80 &\textbf{0.06} &0.88 &0.88 &0.00 \\
		BUSBRA   &0.63 &0.65 &0.02 &0.48 &0.66 &\textbf{0.18} &0.61 &0.68 &\textbf{0.07} &0.85 &0.80 &\textbf{-0.05} &0.54 &0.64 &\textbf{0.10} &0.71 &0.74 &0.03 \\
		BUSI     &0.67 &0.67 &0.00 &0.50 &0.65 &\textbf{0.15} &0.63 &0.68 &\textbf{0.05} &0.67 &0.68 &0.01 &0.69 &0.68 &-0.01 &0.66 &0.67 &0.01 \\
		BUSI\_WHU&0.71 &0.70 &-0.01 &0.58 &0.78 &\textbf{0.20} &0.67 &0.73 &\textbf{0.06} &0.77 &0.81 &0.04 &0.61 &0.67 &\textbf{0.06} &0.83 &0.82 &-0.01 \\
		\hline
		Mean&0.74 &0.73 &-0.01 &0.57 &0.72 &\textbf{0.15} &0.70 &0.75 &\textbf{0.05} &0.77 &0.78&0.01 &0.63 &0.68 &\textbf{0.05} &0.75 &0.76 &0.01 \\
		\hline
	\end{tabular}
	\caption{Recall values for all model-dataset-pairs. The row names represent the datasets used for evaluation, while the column names indicate the dataset each model was trained on. The columns "Ori," "PCA," and "Diff" represent the original model, the model trained on PCA-reconstructed images, and the difference in performance, respectively. Bold values highlight substantial improvements (Diff $\geq 0.05$) and deteriorations (Diff $\leq -0.05$).}
\end{table*}

\begin{figure}[]
	\centering
	\includegraphics[width=\columnwidth]{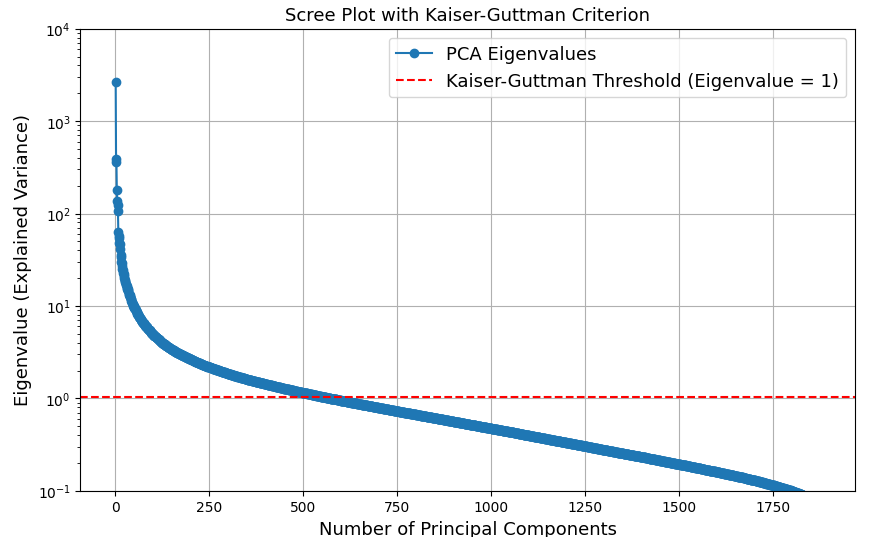} 
	\caption{Exemplary scree plot of the BUSBRA dataset showing the eigenvalues of the principal components. The Kaiser-Guttman criterion (eigenvalue > 1) was applied to determine the optimal number of components to retain for dimensionality reduction.}
	\label{fig:ScreePlot}
\end{figure}

By analyzing scree plots and applying the Kaiser-Guttman criterion~\cite{PCA}, we determined the optimal number of principal components for each dataset. Figure 2 presents an exemplary scree plot of the BUSBRA dataset, illustrating how the Kaiser-Guttman criterion was used to identify components with eigenvalues greater than one. To confirm that the selected number of components accounts for approximately 90\% of the total variance, we examined the cumulative explained variance (Fig. 3). This analysis ensured that the chosen components retained the majority of the data signal while discarding noise.

This selection process was applied to each dataset, balancing the retention of critical information with noise reduction, consistent with best practices in dimensionality reduction. Retaining these components aims to enhance segmentation performance by focusing on the most informative features, reducing overfitting, and improving generalizability across diverse datasets.
\begin{figure}[]
	\centering
	\includegraphics[width=\columnwidth]{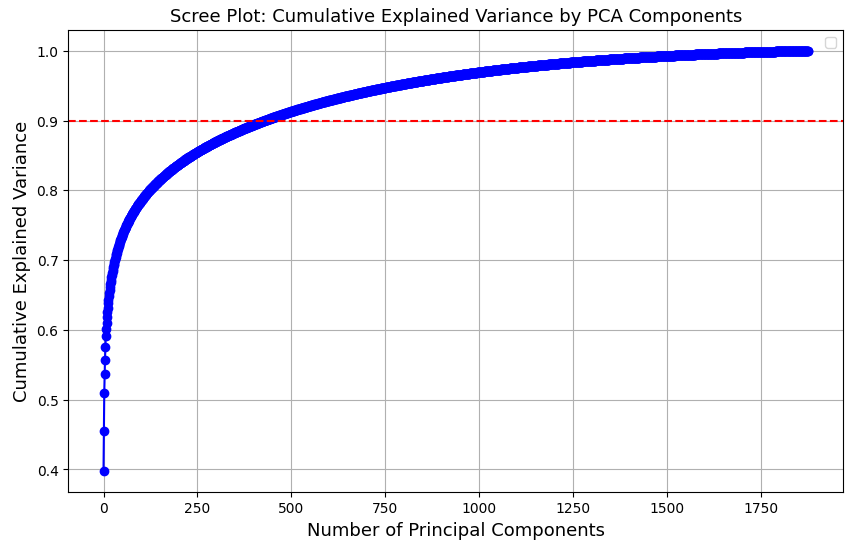} 
	\caption{Cumulative explained variance graph for the BUSBRA dataset, confirming that the number of principal components selected from the scree plot accounts for approximately 90\% of the total variance.}
	\label{fig:ExplainedVariance}
\end{figure}

\setlength{\tabcolsep}{2pt} 
\begin{table*}[t]
	\centering
	\begin{tabular}{|c|ccc|ccc|}
		\hline
		& \multicolumn{3}{c|}{Recall} & \multicolumn{3}{c|}{Dice} \\
		\cline{2-7}
		& Ori & PCA & $\mathit{p}$ & Ori & PCA & $\mathit{p}$ \\
		\hline
		All pairs &$0.68 \pm 0.10$ &$0.72 \pm 0.07$ & 0.08 &$0.59 \pm 0.11$ &$0.64 \pm 0.10$ &0.16 \\
		Worst 10  &$0.57 \pm 0.07$ &$0.70 \pm 0.05$ &\textbf{0.0004} &$0.50 \pm 0.06$ &$0.58 \pm 0.06$ & \textbf{0.03}\\
		Other 20  &$0.73 \pm 0.08$ &$0.74 \pm 0.08$ &0.94 &$0.64 \pm 0.10$ &$0.66 \pm 0.10$ &0.62 \\
		\hline
	\end{tabular}
	\caption{Comparison of segmentation results of original and PCA images (Recall, Dice, and $\mathit{p}$-values). "Worst 10" refers to the worst 10 model-dataset-pairs in terms of external recall-decline when using original images, while "Other 20" refers to the rest of the 30 model-dataset-pairs.}
\end{table*}

To evaluate the impact of PCA on external validity, twelve models were trained: six on the original datasets and six on the PCA-reconstructed datasets. Each model trained on an original dataset was validated on the remaining five out-of-domain original datasets, while each model trained on a PCA dataset was validated on five out-of-domain PCA datasets. This produced a total of 60 model-dataset pairs, 30 pairs using the original datasets and 30 pairs using the PCA datasets. Comparisons were then made between the performances of models trained on the original versus PCA datasets. Model performances were compared using two-tailed $t$-tests and a significance level of $\alpha=0.05$.

\section{Results}

Our primary performance evaluation metric was the recall value, given the substantial consequences and health implications associated with misidentifying tumor pixels as false-negative. Additionally, we kept track of the Dice score to monitor the overall segmentation quality. The summarized results (Table 2) present the segmentation recall scores for each model-dataset pair, along with any change observed from using models trained with PCA-reconstructed images. The comparison between using the original and PCA datasets for model training shows substantial mean improvements $\geq 0.05$ in 3 out of 6 models. For all model-dataset pairs (Table 3), PCA led to a moderate improvement in recall (0.72 $\pm$ 0.07) compared to the original dataset (0.68 $\pm$ 0.10), with a \textit{p}-value of 0.08, which indicates moderate evidence for an effect, though it does not meet the threshold for statistical significance. The improvement in Dice score, though not statistically significant (\textit{p} = 0.16), also showed a positive trend, with the PCA dataset achieving a score of 0.64 $\pm$ 0.10 compared to 0.59 $\pm$ 0.11 for the original dataset. On average, the absolute mean decline in recall values during external validation of initially 0.12 was reduced by 33\% to 0.08. Notably, when focusing on the 10 worst-performing pairs, identified by the largest decrease in recall during external validation, PCA provided a substantial improvement. Recall increased from 0.57 $\pm$ 0.07 in the original dataset to 0.70 $\pm$ 0.05 with PCA (\textit{p} = 0.0004), and the Dice score rose from 0.50 $\pm$ 0.06 to 0.58 $\pm$ 0.06 (\textit{p} = 0.03), both statistically significant improvements. 

The visual segmentation results (Fig. 4) demonstrate multiple types of improvements achieved by the PCA-preprocessed models. In several cases, the PCA model effectively removes false positive background artifacts, resulting in cleaner segmentation outputs. Additionally, the PCA model enhances the geometric accuracy of tumor shapes, producing more anatomically plausible segmentations compared to the original model. Notably, in some challenging instances where the original model failed to detect tumors entirely, the PCA model successfully identifies and segments these regions with a reasonable degree of accuracy.

\section{Discussion}

The results in Table 3 indicate that the proposed use of PCA reconstruction serves as a safeguard in challenging cases where external validity is low, helping to protect against sharp declines in segmentation accuracy. This is likely caused by reducing noise and emphasizing key features relevant to image segmentation. This conclusion is supported by the significant improvements in both recall and Dice scores observed in the worst-performing model-dataset pairs, where external validation showed the largest drop in performance. However, for the other 20 model-dataset pairs, PCA reconstruction did not lead to a significant improvement in either recall (\textit{p} = 0.94) or Dice scores (\textit{p} = 0.62). This suggests that the primary impact of the PCA reconstruction lies in enhancing performance when the model struggles with external validity. On the other hand, PCA offers only limited additional benefits for external cases with already well-performing image segmentation. 

It's worth noting that the effectiveness of PCA varies significantly across datasets, prompting the question of why some datasets benefit from this method much more than others. Another interesting observation is the disproportionate improvement in recall compared to Dice scores. This could be explained by the fact that PCA blurs the margins of tumors, making them appear more diffuse and thus easier to detect in terms of overall presence, yet at the cost of boundary precision.

Comparing our findings to those of Yu et al~\cite{Yu}, it's interesting to note that while the authors report a decrease in external performance in radiologic diagnosis (in terms of recall/precision) larger than 0.05 for half of their reviewed studies, that was the case for about two-thirds of our 30 model/dataset pairs. This might be attributed to a larger inter-operator and inter-device variability in our ultrasound application, compared to their radiographic images.

The visual segmentation results (Fig. 4) highlight key improvements achieved by PCA-preprocessed models, including the reduction of false positive background artifacts and enhanced geometric accuracy of tumor shapes. These refinements suggest that PCA effectively filters out noise while preserving critical structural information. Furthermore, the PCA models demonstrated the ability to detect tumors that were entirely missed by the original models, showcasing their potential to enhance sensitivity in challenging cases. 

For future work, it would be valuable to conduct ablation studies on the effects of varying the number of principal components to optimize internal consistency, external validity, and model complexity. Expanding these studies to a larger set of datasets could reveal how the proportion of principal components impacts metrics such as precision and recall, offering deeper insights into fine-tuning PCA for improved generalizability across diverse datasets. Furthermore, combining PCA with established methods and integrating PCA in existing approaches, such as data augmentation or style transfer, may result in further improvements to external validity. While augmentation is a standard practice to enhance generalization, it was excluded in this work to isolate and evaluate the specific impact of PCA preprocessing on external validity. In addition, future studies could explore the comparative effectiveness of PCA against traditional filtering techniques, evaluate the impact of label variability across datasets on segmentation performance, and incorporate additional metrics such as AUROC or boundary distances to provide a more comprehensive assessment of model generalization and segmentation quality. While our experiments were limited to a basic U-Net architecture, the results serve as a proof of concept and future work should evaluate the impact of PCA preprocessing across more advanced segmentation models to assess the generalizability of these findings. Our results indicate that PCA may be particularly useful in cases where recall is prioritized over precision. Another hypothesis to be further investigated is whether datasets requiring more principal components to achieve a certain percentage of explained variance capture more complex or nuanced features, potentially leading to further enhancement of model generalizability.

\section{Acknowledgment}
This work was funded by the German Federal Ministry of Education and Research (BMBF) under the program KMU-innovativ: Medizintechnik (project name: MammaSound, grant number 13GW0703B).

\begin{figure*}[b]
	\centering
	\includegraphics[width=\textwidth]{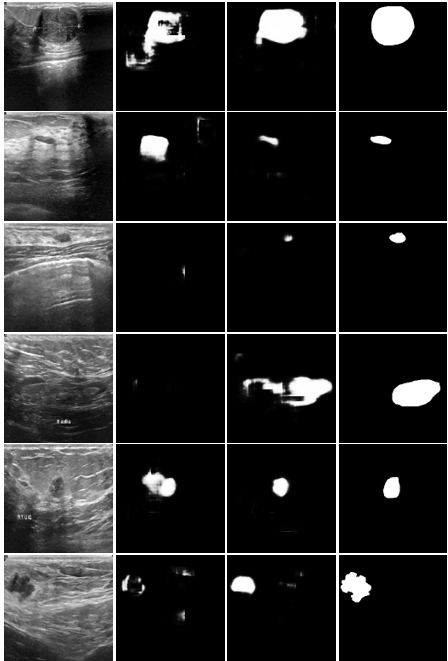} 
	\caption{Comparison of exemplary visual segmentation results for the model-dataset-pair BUSI-BUS\_UC. First column: original US image, second column: segmentation result using orignal dataset model, third column: segmentation result using PCA model (our method), fourth column: ground truth segmentations from the BUSI dataset }
	\label{fig:Segmentierung}
\end{figure*}

\end{document}